\documentclass[aps,prx,showpacs,floatfix,onecolumn,superscriptaddress,hyperref]{revtex4-2}

\usepackage{epsfig,graphicx,amsmath,amssymb,color}
\usepackage{subcaption}
\usepackage{hyperref}
\usepackage{amsfonts}
\usepackage{mathrsfs}
\usepackage{mathtools}
\usepackage{braket}
\usepackage{physics}

\newcommand{\new}[1]{{\color{black}{#1}}}
\newcommand{\newnew}[1]{{\color{black}{#1}}}

\begin{document}

\title{Experimental Test of Nonlocality Limits from Relativistic Independence}

\author{Francesco Atzori}
\affiliation{INRIM, Strada delle Cacce 91, I-10135 Torino, Italy}
\affiliation{Politecnico di Torino, Corso Duca degli Abruzzi 24, I-10129 Torino, Italy}

\author{Salvatore Virz\`{i}}
\author{Enrico Rebufello}
\author{Alessio Avella}
\affiliation{INRIM, Strada delle Cacce 91, I-10135 Torino, Italy}

\author{Fabrizio Piacentini}
\email{f.piacentini@inrim.it}
\affiliation{INRIM, Strada delle Cacce 91, I-10135 Torino, Italy}

\author{Iris Cusini}
\author{Henri Haka}
\author{Federica Villa}
\affiliation{Politecnico di Milano, Dipartimento di Elettronica, Informazione e Bioingegneria, Piazza Leonardo da Vinci 32, 20133 Milano, Italy}

\author{Marco Gramegna}
\affiliation{INRIM, Strada delle Cacce 91, I-10135 Torino, Italy}

\author{Eliahu Cohen}
\affiliation{Faculty of Engineering and the Institute of Nanotechnology and Advanced Materials, Bar Ilan University, Ramat Gan, Israel}

\author{Ivo Pietro Degiovanni}
\author{Marco Genovese}
\affiliation{INRIM, Strada delle Cacce 91, I-10135 Torino, Italy}
\affiliation{INFN, sezione di Torino, via P. Giuria 1, 10125 Torino, Italy}


\baselineskip24pt


\begin{abstract}
Quantum correlations, like entanglement, represent the characteristic trait of quantum mechanics, and pose essential issues and challenges to the interpretation of this pillar of modern physics.
Although quantum correlations are largely acknowledged as a major resource to achieve quantum advantage in many tasks of quantum technologies, their full quantitative description and the axiomatic basis underlying them are still under investigation.
Previous works suggested that the origin of nonlocal correlations is grounded in principles capturing (from outside the quantum formalism) the essence of quantum uncertainty.
In particular, the recently-introduced principle of Relativistic Independence gave rise to a new bound intertwining local and nonlocal correlations.\\
Here we test such a bound by realizing together sequential and joint weak measurements on entangled photon pairs, allowing to simultaneously quantify both local and nonlocal correlations by measuring incompatible observables \newnew{on the same quantum system without collapsing its state,} a task typically forbidden in the traditional (projective) quantum measurement framework.
Our results demonstrate the existence of a fundamental limit on the extent of quantum correlations, shedding light on the profound role of uncertainty in both enabling and balancing them.
\end{abstract}
\maketitle

\section{Introduction}

Investigating the very principles behind the laws of nature is one of the most captivating and intricate fields of scientific research.
In this sense, quantum mechanics (QM) has demonstrated extraordinary predictive power when modeling the behavior of microscopic particles, providing also tools for technological boosts in fields like communication \cite{luc18,xu20,pir20,deq21,che21,cli22}, computation \cite{zho20,mad22,dal22,con22,kim23}, imaging \cite{ruo10,bri11,gen16,med17,mor19,cas21}, hypothesis testing \cite{ort21a,var21,ort21b,kar24}, metrology \cite{gio11,pez18,bar22,fal22,yin23} and sensing \cite{aha14,bern17,deg17,pir18,oli18,pet22,vir22,vir24}, but a large debate is still running in the scientific community about its foundational aspects.
Particularly relevant is the investigation of nonlocal traits of quantum correlations among spatially-separated parties \cite{bru13}, a research field which originated in 1935 with the Einstein-Podolsky-Rosen (EPR) paradox \cite{epr}.
In 1964, J. S. Bell demonstrated how locality constrains the correlations between measurements on a bipartite system \cite{bel65}, proving that QM can exceed such a bound and, therefore, proving that its predictions are incompatible with those of any (classical) Local Hidden Variable Theory (LHVT) \cite{bel65,gen05}.\\
Later on, B. S. Tsirelson made an important step towards a quantitative analysis of quantum mechanical correlations, demonstrating that they are nevertheless bounded \cite{tsi80}.
Subsequently, many physicists put large efforts towards the investigation of quantum correlations and the first principles determining their strength \cite{lin07,paw09,opp10,nav10,dam13,fri13,hof19}, also trying to rule out probabilistic models exhibiting stronger-than-quantum correlations without violating the no-signaling principle \cite{pop94}.
However, none of these attempts managed to account for the whole set of one- and two-point correlators in the simplest bipartite two-outcome scenario.\\
Recently, A. Carmi \& E. Cohen proposed a new fundamental principle, called Relativistic Independence (RI) \cite{car19}, stating that generalized local uncertainty relations (most generally defined, even outside quantum mechanics) cannot be affected by space-like separated parties.
Similarly to the works in Refs. \cite{opp10,hof19}, this approach emphasizes the major role of uncertainty in nature \cite{coh20}.
In addition, it allows deriving familiar as well as novel bounds on quantum correlations, without explicitly assuming the full mathematical structure of QM.
Indeed, correlations arising in theories complying with RI must satisfy certain bounds displaying an interplay between the nonlocal correlations shared among the parties of a multipartite system and the local correlations within the lab of each party (e.g., those related to quantum uncertainty) \cite{car19,car18,coh20}, ruling out theories admitting stronger-than-quantum correlations.
Such RI-derived bounds generalize other known bounds on nonlocal quantum correlations, and their experimental study is the main topic of this work.\\
The need to estimate both local and nonlocal correlations at once makes the experimental test of these bounds quite a challenging task, requiring to quantify correlations among several incompatible (non-commuting) observables measured \newnew{on the same quantum system without radically changing its state}, which is prohibited by Heisenberg's uncertainty principle and by the wave function collapse in the standard projective measurement scheme.
In the following, we show how we managed to test the RI bound by producing polarization-entangled photon pairs and performing a sequence of two weak measurements (WMs) \cite{aha88,rit91,dre14,lun09,rin14,cal20,mar21,mit07,the16,pia16seq,ave17,pia18,kim18,fol21} on both photons of each pair, allowing to quantify all the correlations needed for testing the RI bound.
Our results lead to a new experimental characterization of the set of nonlocal correlations within entangled states while emphasizing the role of local correlations in determining and bounding them.
We also demonstrate the operative meaning of the bound stemming from RI, thereby clarifying its physical interpretation.\\
\section{Theoretical framework}
The correlations arising in quantum theory are much more intricate and interesting than those appearing in classical physics.
In particular, because of quantum entanglement, measurements made on distant systems can lead to nonlocal correlations, i.e., correlations that are classically inexplicable because of their incompatibility with local realism (provided that the ``free choice assumption'' or ``statistical independence'' holds \cite{bel16,ber17,myr21}).
This statement is captured by the violation of Bell inequalities \cite{bru13,gen05}, one of which is the Clauser-Horne-Shimony-Holt (CHSH) inequality:
\begin{equation}
 \mathcal{|B|} \equiv \left|\langle A_1B_1\rangle +\langle A_1B_2\rangle+\langle A_2B_1\rangle-\langle A_2B_2\rangle\right| \le 2,
\end{equation}
where $\mathcal{B}$ is the so-called Bell-CHSH parameter and $A_i$,$B_j$ are local dichotomic variables measured in the labs of two observers, Alice ($A$) and Bob ($B$), with two different measurement settings, i.e., $i,j=1,2$.
In quantum mechanics they are represented by Hermitian operators, whose nonlocal correlations may exceed the value of 2 reaching the aforementioned Tsirelson bound, i.e.:
\begin{equation}
\mathcal{|B|} \le 2\sqrt{2},
\end{equation}
which determines the maximal extent of nonlocal quantum correlations.
This inequality sets a bound on quantum correlations in bipartite scenarios without specifying how it is approached, i.e. how the bound on non-maximal quantum correlations looks like and what makes them non-maximal.
Subsequently, richer bounds were proposed \cite{uff02,tsi87,lan88,mas03,cab04,fil04,weh06,nav07,goh18} and tested \cite{chr15}.\\
In this work, we seek to advance the frontiers of this expedition towards the ultimate set of quantum correlations, exploring novel and more elaborated bounds on nonlocal (quantum) correlations.
Remarkably, we experimentally show that these bounds on nonlocal correlations between Alice and Bob as captured by $\mathcal{B}$, stem from the local correlations between their operators $A_1$, $A_2$ on one side, and $B_1$,$B_2$ on the other.
Moreover, they prove that a necessary condition for quantum correlations to achieve their maximum value is zero correlation between the two measurements on each party's lab.

\subsection*{Generalized uncertainty relations and the Relativistic Independence bound}
\newnew{The relativistic independence (RI) principle \cite{car19} stems from an attempt to quantify the strength of quantum correlations from outside the quantum formalism.
It therefore aims at encoding two basic requirements, uncertainty and locality (in the sense that local uncertainty relations do not depend on measurement choices made by other parties), using a general statistical structure, namely the covariance matrix, to which only the measured outcomes can be inserted (rather than other theoretical constructs such as Hermitian operators, commutators, etc.).
The first point to be recalled here is the affinity between the Robertson-Schr\"odinger uncertainty relations and the semidefinite-positiveness of the corresponding covariance matrix \cite{car18,coh20} - this will be illustrated in Eq. \eqref{genrel} below.
The second point, which may require some conceptual leap, is the ability to transcend single-system uncertainty relations to bi- (or in general multi-) partite uncertainty relations.
Finally, the RI principle ensures that such multipartite uncertainty relations encoded within covariance matrices will depend only on local parameters and, from that, it enables to derive various bounds on quantum correlations.
Then, let our two observers, Alice and Bob, share a bipartite system.}
As stated above, they can measure the physical variables $A_i$, $B_j$ and they can estimate the variances $\Delta_{A_i}^2$ and $\Delta_{B_j}^2$ and the covariances $C(B_j,A_i) = E_{B_j A_i} - E_{B_j}E_{A_i}$, where $E_{A_i}$, $E_{B_j}$ and $E_{B_jA_i}$ are the one- and two-point correlators, respectively.
The full system is governed by a generalized uncertainty relation which can be written as the following positive semi-definiteness condition \cite{car19} (see Appendix A):\newnew{
\begin{equation}
\pmb{\Lambda}_{AB}=
\begin{bmatrix}
\pmb{\Lambda}_B 		& \pmb{C} (B,A) \\
\pmb{C}^\dagger (B,A)	 	& \pmb{\Lambda}_A	
\end{bmatrix}\succeq 0,
\end{equation}
where $\pmb{C} (B,A)$ is the cross-covariance matrix \cite{gub06} between Alice and Bob, costituted by the $C(B_j,A_i)$ elements and representing the nonlocal correlations of the bipartite system, while $\pmb{\Lambda}_A$ and $\pmb{\Lambda}_B$ are the local covariance matrices for $A$ and $B$, respectively, representing the local uncertainty relations of each subsystem, i.e. the familiar ones pertaining to $A_1,A_2$ in Alice's lab and $B_1,B_2$ in Bob's lab.
Now, let Bob measure just one variable $B_j$, choosing one of the two available measurement settings.
The generalized uncertainty relations are obtained by imposing positive semi-definiteness to a submatrix of $\pmb{\Lambda}_{AB}$, i.e. the existence of some $r_j$ from which stems the expression:}
\begin{equation}
\pmb{\Lambda}_{AB}^j=
\begin{bmatrix}
\Delta_{B_j}^2		& C(B_j,A_1)		&C(B_j,A_2) \\
C^*(B_j,A_1) 		&\Delta_{A_1}^2 	&r_j \\
C^*(B_j,A_2)		&r^*_j			& \Delta_{A_2}^2	
\end{bmatrix}\succeq 0.
\label{genrel}
\end{equation}
LHVTs, QM and even some models allowing for stronger-than-quantum correlations all satisfy Eq. \eqref{genrel}.
The $r_j$ term is typically unmeasurable, as it corresponds to the non-observable correlations between local variables, and its form depends on the considered theory; in QM, e.g., one has $r_1=r_2\equiv r^Q=\frac{\langle \hat{A}_1 \hat{A}_2 + \hat{A}_2 \hat{A}_1 \rangle}{2} - \langle \hat{A}_1\rangle\langle \hat{A}_2\rangle$.
In this case, Eq. \eqref{genrel} leads to the Schr\"{o}dinger-Robertson uncertainty relation applied to the nonlocal operators $B_jA_1$ and $B_jA_2$ \cite{car19}.
Unfortunately, the uncertainty principle and wave function collapse do not allow measuring $r^Q$ together with all other quantities appearing in Eq. \eqref{genrel}, at least within the traditional (projective) quantum measurement framework.\\
The fact that the $r$ coefficient might depend on $j$ means that, in principle, even in a space-like separated scenario Bob might still have a (nonlocal) influence on the generalized uncertainty relations.
The RI principle forbids such nonlocal influence, stating that, in case of space-like separation, the generalized uncertainty relations on Alice's (Bob's) side are independent of what occurs in Bob's (Alice's) lab, i.e.:
\begin{equation}
\pmb{\Lambda}_{A}^j=
\begin{bmatrix}
\Delta_{A_1}^2 	&r_j \\
r^*_j		& \Delta_{A_2}^2
\end{bmatrix}  \;\;\;\;\stackrel{\mathrm{RI}}{\longrightarrow}\;\;\;\; \pmb{\Lambda}_{A} =
\begin{bmatrix}
\Delta_{A_1}^2 	&r \\
r^*		& \Delta_{A_2}^2
\end{bmatrix}.
\label{RImat}
\end{equation}
From this formulation, one can derive the following bound (see Appendix A), displaying the interplay between local correlations within each party's lab and the nonlocal correlations arising between the two parties:\newnew{
\begin{equation}
0\leq\mathcal{RI}=\left|\frac{\mathcal{B}}{2\sqrt{2}}\right|^2 + \left(\Re\left[ \frac{r}{2\Delta_{A_2}\Delta_{A_1}} \right]\right)^2 \leq 1\;.
\label{RI}
\end{equation}
In the QM case, i.e. for $r=r^Q$, Eq. \eqref{RI} becomes
\begin{equation}
	0\leq\mathcal{RI}=\abs{\frac{\mathcal{B}}{2\sqrt{2}}}^2 + \Delta^2 \leq 1\;,
%
%
	\label{eq:RI-split}
\end{equation}
where $\mathcal{B}$ represents the nonlocal correlations contribution and $\Delta=\frac{r^Q}{2\Delta_{A_2}\Delta_{A_1}}$, being proportional to the real part of the Pearson correlation coefficient $\frac{\langle A_2 A_1 \rangle-\langle A_2 \rangle\langle A_1 \rangle}{\Delta_{A_2}\Delta_{A_1}}$ \cite{maccons}} between the two measurements occurring in Alice's lab, accounts for the (local) correlations stemming from the uncertainty relations on Alice's side.\\
\newnew{Eqs. \eqref{RI} and \eqref{eq:RI-split} suggest that in quantum mechanics, but also more generally, large nonlocal correlations and large local correlations cannot co-exist.
Put differently, to have violations of the CHSH inequality, we must have small correlation between the local choices of Alice (and similarly for Bob).}
%

\section{The experiment}
\new{To test the RI bound in Eq. \eqref{eq:RI-split}, it is required to simultaneously evaluate the (nonlocal) correlations between $A$ and $B$ measurements and the (local) correlations between the two measurements performed either in $A$ or $B$ labs.
This corresponds to evaluating, at once, the Bell-CHSH parameter $\mathcal B$ as well as the Pearson's correlator between the two measurements realized in one of the labs, which we choose to be Alice's.
In general, these measurements involve incompatible (non-commuting) observables, resulting in this task being forbidden by the wave function collapse and Heisenberg's uncertainty principle in the traditional quantum measurement framework based on projective measurements.
However, such a condition can be relaxed by implementing a sequence of two WMs per branch, as shown in Fig. \ref{scheme}, thus avoiding the wave function collapse (together with the inevitable entanglement breaking) at the cost of some tiny decoherence affecting the entangled state after the measurement process (see Appendices B and C for details).
This way, each entangled pair undergoes all the measurements (i.e., two per photon) needed to evaluate the local and nonlocal correlations forming the RI bound.\\
\begin{figure}[htbp]
\includegraphics[width=1.0\columnwidth]{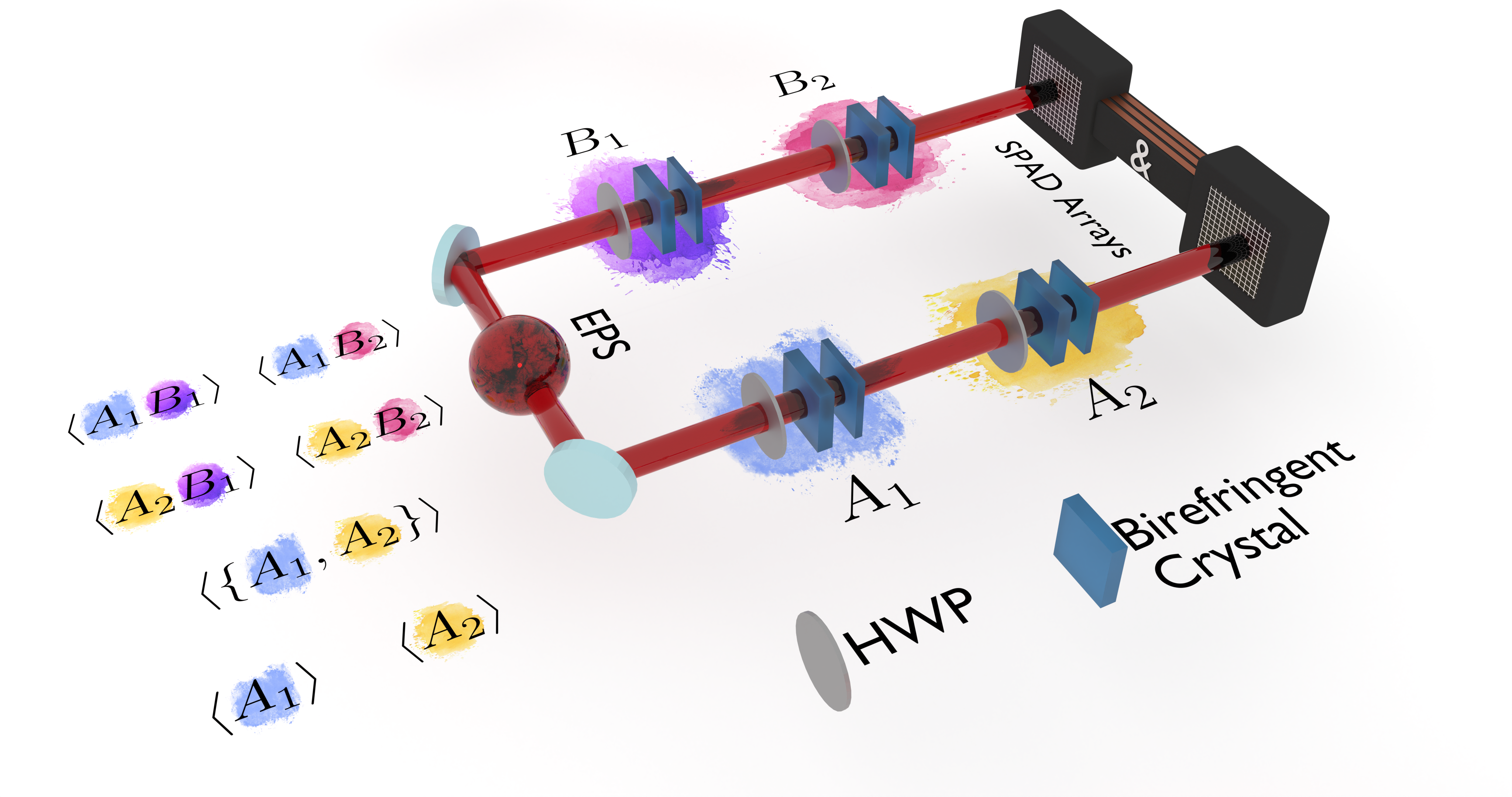}
\caption{\textbf{Scheme of the experimental setup}.
Our entangled pair source (EPS), based on Spontaneous Parametric Down-Conversion (SPDC) occurring in a Sagnac interferometer, generates entangled photon pairs in the state $\ket{\psi_-}=(\ket{H_AV_B}-\ket{V_AH_B})/\sqrt{2}$, sending one photon to Alice ($A$) and the other to Bob ($B$).
Both $A$ and $B$ implement two WMs in a row, each carried by a pair of birefringent crystals preceded by a half-wave plate (HWP), setting the measurement bases $i,j=1,2$.
This way, our two-photon state undergoes joint \cite{lun09,rin14,kum17,cal20,mar21} and sequential \cite{mit07,the16,pia16seq,ave17,kim18} WMs at once, preventing the $\ket{\psi_-}$ state wave function from collapsing although at the cost of some small decoherence, allowing to estimate all the measurement correlations (highlighted by different colors, with $\{A_1,A_2\}=(A_1A_2+A_2A_1$) required for testing the RI bound in Eq. \eqref{eq:RI-split}.
Finally, both $A$ and $B$ photons are detected by a 2D spatial-resolving detector with internal time-tagging capability.
A detailed description of the experimental setup is reported in Appendix B.}
\label{scheme}
\end{figure}
\newnew{Specifically, our entangled pair source (EPS) generates polarization-entangled photon pairs in the singlet state $|\psi_-\rangle=\frac{1}{\sqrt2}\left(|H_AV_B\rangle-|V_AH_B\rangle\right)$ ($H$ and $V$ indicating, respectively, the horizontal and vertical polarization components) via degenerate Spontaneous Parametric Down-Conversion (SPDC) in a Sagnac interferometer.
Both photons of each pair are spectrally filtered by narrow-band interference filters and then spatially filtered by coupling them into single-mode optical fibers, getting decoupled and collimated into Gaussian spatial distributions before being addressed to Alice and Bob (see Appendix B).
This results in an overall initial wave function of our two-photon state of the form $\ket{\Psi_{in}}= \ket{\psi_-} \otimes \ket{f_{x_A}} \otimes \ket{f_{y_A}} \otimes\ket{f_{x_B}} \otimes\ket{f_{y_B}}$, being $\bra{\zeta}\ket{f_\zeta} =\frac{1}{(2\pi\sigma^2)^{1/4}}\exp\left(-\frac{\zeta^2}{4\sigma^2}\right)$ the $\sigma$-wide Gaussian distribution of the photons in each transverse direction $\zeta=x_A,y_A,x_B,y_B$.
The generated singlet state presents visibility $V^{in}=\frac{N(+,-)+N(-,+)-N(+,+)-N(-,-)}{N(+,-)+N(-,+)+N(+,+)+N(-,-)}=0.983\pm 0.001$, being $N(\pm,\pm)$ the number of two-photon counts registered while projecting photons $A$ and $B$ onto the states defining the diagonal/antidiagonal polarization basis, i.e. $\ket{\pm}=\frac{1}{\sqrt2}\left(\ket{H}\pm\ket{V}\right)$.}\\
In both Alice's and Bob's labs, each WM is realized by exploiting the weak coupling between polarization and transverse momentum induced on the photon by a pair of thin birefringent crystals. 
The first crystal of the pair is the principal one, and it induces a spatial walk-off (for the polarization component lying in the same plane as the crystal extraordinary ($e$) optical axis) in one of the two independent transverse directions $x$ (for the first WM) and $y$ (for the second one), leaving the two WMs occurring on each photon independent of each other.
Anyway, this crystal also introduces a temporal delay and a phase shift between the $H$ and $V$ polarization components; the following ``compensation'' crystal, with the optical $e$-axis oriented along the $y$ (for the first WM) and $x$ (for the second one), is rotated along its optical $e$-axis to compensate the temporal delay and nullify the phase shift without introducing any spatial decoherence.
Each crystal pair is preceded by a half-wave plate (HWP), allowing to select the polarization basis of the subsequent WM; the combined effect of HWP and crystal pair realizes the unitary transformation:
\begin{equation}
\hat{U}_{\zeta_{j K}} =\exp \left( -\frac{i}{\hbar}g_{\zeta_{j K}}\hat{\Pi}(\vartheta_{K_j} )\otimes \hat{P}_{\zeta_{j K}} \right),
\label{Ugen}
\end{equation}
where $g_{\zeta_{j K}}$ is the coupling constant ($K=A,B$, $j=1,2$, $\zeta_1=x$, $\zeta_2=y$) indicating the measurement strength, $\hat{\Pi}(\vartheta_{K_j})$ is the polarization projector along the direction described by the angle $\vartheta_{K_j}$, and $\hat{P}_{\zeta_{j K}}$ is the transverse momentum of the photon on the optical plane of the crystal.
\newnew{For the WM condition to hold, the coupling constants $g_{\zeta_{j K}}$ must be much smaller than the spread of the pointer observable \cite{aha88}, i.e., in the specific case, of the width $\sigma$ of the spatial Gaussian distribution of the photons (a thorough study of the measurement strength $g$ effect on the WM can be found in \cite{pia18}).
This is needed to grant an (almost) negligible back-action of the measurement on the quantum system undergoing it, thus preventing the wave function collapse.
Although, in principle, this would imply $\frac{g_{\zeta_{j K}}}{\sigma}\ll1$, it has been shown \cite{pia18} that, for the kind of measurements needed in our experiment, it is possible to achieve such a regime already for $\frac{g_{\zeta_{j K}}}{\sigma}\lesssim0.2$, as implemented in our setup (see Appendix C for details).}\\
After the four WMs, the entangled photon pairs are addressed to two spatially-resolving detectors, allowing to extract for each two-fold coincidence the coordinates of the firing pixels and thus obtaining the coincidence counts tensor $N(X_A,Y_A,X_B,Y_B)$, being $X_A$, $Y_A$ and $X_B$, $Y_B$ the pixel coordinates in which the two photon of each pair impinge.\\
By relying just on WMs, i.e. without any postselection added after the weak couplings, we avoid the wave function collapse and manage to recover all the one- and two-point correlators needed for testing the RI bound.
Indeed, considering only projections onto the real plane of the Bloch sphere, after the entangled pair undergoes the four WMs one can extract the cross-correlations:
\begin{equation}
\langle \hat{\zeta}_{jA} \otimes \hat{\zeta}_{lB} \rangle_{out} = \bra{\Psi_{out}} \hat{\zeta}_{jA} \otimes \hat{\zeta}_{lB} \ket{\Psi_{out}} \simeq g_{\zeta_{jA}} g_{\zeta_{lB}} \bra{\psi_-} \hat{\Pi}(\vartheta_{A_j} )\otimes  \hat{\Pi}(\vartheta_{B_l}) \ket{\psi_-},
\label{crosscorr}
\end{equation}
%
the single-branch sequential correlations:
\begin{equation}
\langle \hat{X}_{K} \hat{Y}_{K} \rangle_{out} = \bra{\Psi_{out}} \hat{X}_{K} \hat{Y}_{K} \ket{\Psi_{out}} \simeq \frac{g_{x_{K}} g_{y_{K}}}{2} \bra{\psi_-} \left\{\hat{\Pi}(\vartheta_{K_1} ),\hat{\Pi}(\vartheta_{K_2})\right\} \ket{\psi_-},
\label{seqcorr}
\end{equation}
and the single-observable expectation values:
\begin{equation}
\langle \hat{\zeta}_{jK}\rangle_{out} = \bra{\Psi_{out}} \hat{\zeta}_{jK}\ket{\Psi_{out}} \simeq g_{\zeta_{jK}} \bra{\psi_-} \hat{\Pi}(\vartheta_{K_j} )\ket{\psi_-},
\label{1point}
\end{equation}
where $j,l=1,2$ and $K= A,B$, the projectors $\hat{\Pi}(\vartheta_{K_j})$ are defined as $\hat{\Pi}(\vartheta_{K_j})= \frac{\hat{I} + \hat{\sigma}_z (\vartheta_{K_j})}{2}$, the symbol $\{,\}$ indicates the anticommutator, and $\ket{\Psi_{out}}=\hat{U}_{y_A}\hat{U}_{x_A}\hat{U}_{y_B}\hat{U}_{x_B} \ket{\Psi_{in}}$ is the bipartite state after all weak couplings \newnew{(for the sake of readability, we omit from Eq. \eqref{1point} the tensor with the identity operator for the system subspace where no observable is evaluated)}.
Combining all these elements, one can evaluate the $\mathcal{RI}$ quantity in Eq. \eqref{eq:RI-split} as:
\begin{multline}
\mathcal{RI}=\left|\frac{4\left(\frac{\langle \hat{X}_A \hat{X}_B \rangle_{out}}{g_{X_A}g_{X_B}} -  \frac{\langle\hat{X}_A \hat{Y}_B \rangle_{out}}{g_{X_A}g_{Y_B}} + \frac{\langle \hat{Y}_A \hat{X}_B \rangle_{out}}{g_{Y_A}g_{X_B}} +\frac{\langle \hat{Y}_A \hat{Y}_B \rangle_{out}}{g_{Y_A}g_{Y_B}} -\frac{\langle \hat{Y}_A  \rangle_{out}}{g_{Y_A}} -\frac{\langle \hat{X}_B \rangle_{out}}{g_{X_B}} \right)+ 2}{2\sqrt{2}}\right|^2 +\\
+ \left| \frac{\frac{\langle \hat{X}_A \hat{Y}_A \rangle_{out}}{g_{X_A}g_{Y_A}} - \frac{\langle \hat{X}_A \rangle_{out} \langle \hat{Y}_A \rangle_{out}}{g_{X_A}g_{Y_A}}}{2\sqrt{\left( \frac{\langle \hat{X}_A \rangle_{out}}{g_{X_A}}- \frac{\langle \hat{X}_A \rangle_{out}^2}{g^2_{X_A}}\right) \left(\frac{\langle \hat{Y}_A \rangle_{out}}{g_{Y_A}}- \frac{\langle \hat{Y}_A \rangle^2_{out}}{g^2_{Y_A}} \right)}} \right|^2,
\label{RIbound}
\end{multline}
where the first term corresponds to the $\left|\frac{\mathcal B}{2\sqrt2}\right|^2$ value and the second one accounts for the $\Delta^2$ component (a detailed description of the experimental setup can be found in Appendix B).
By comparing Eqs. \eqref{crosscorr}-\eqref{RIbound}, one can appreciate how, in the weak approximation regime (i.e., for small coupling condition $\frac{g_{\zeta_{j K}}}{\sigma} \ll 1$, as in our case), the weakly-measured $\mathcal{RI}$ is independent of the exact $g_{\zeta_{j K}}$ values, since they cancel out when substituting the expressions of Eqs. \eqref{crosscorr}-\eqref{1point} in Eq. \eqref{RIbound}.}

\begin{figure}[t]
	\begin{subfigure}[c]{0.48\textwidth}
		\centering
		\includegraphics[width=\textwidth]{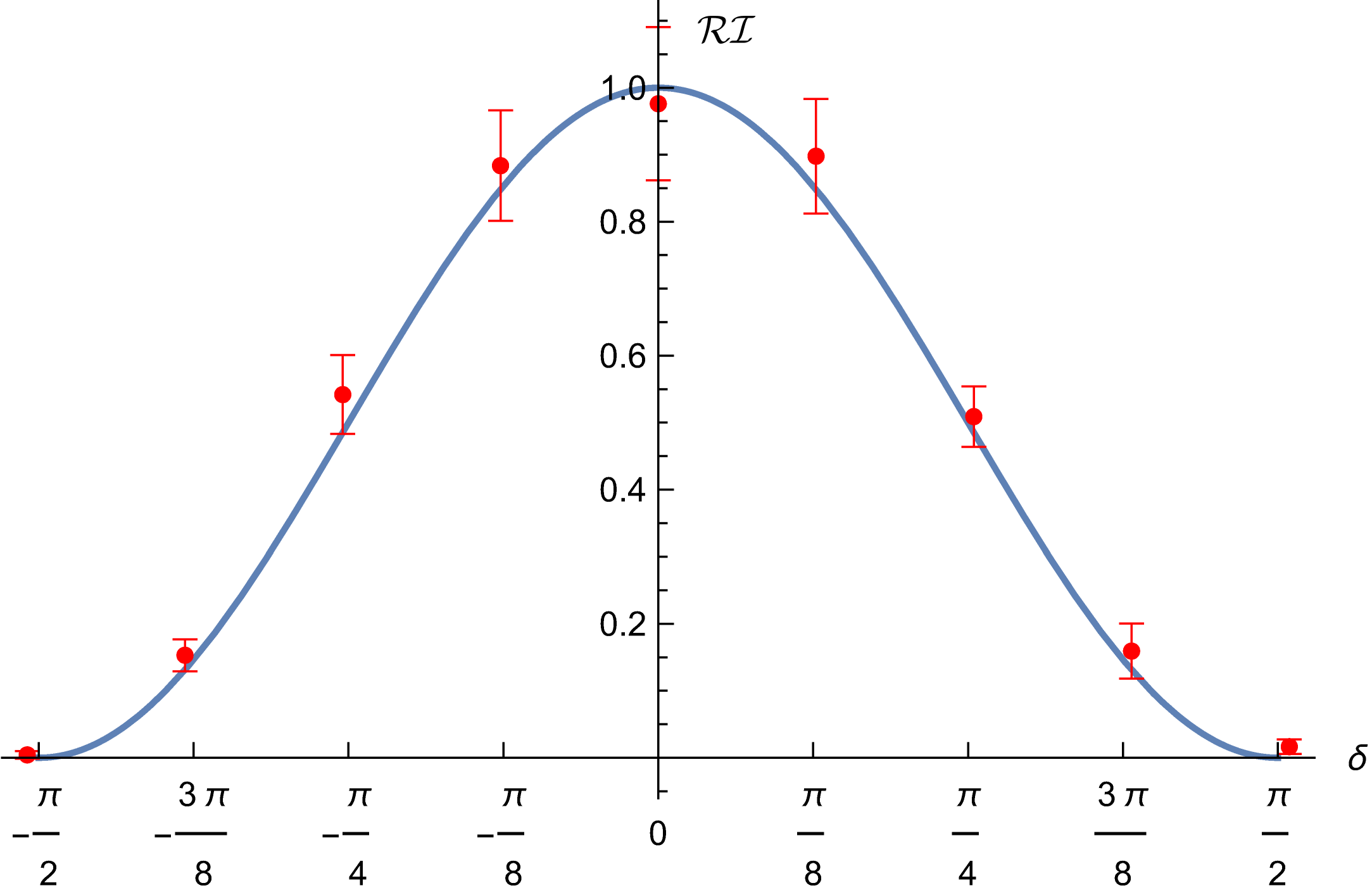}
		\subcaption{Overall $\mathcal{RI}$.}
		\label{totale}
	\end{subfigure}
	\hfill
	\begin{subfigure}[c]{0.48\textwidth}
		\centering
		\includegraphics[width=\textwidth]{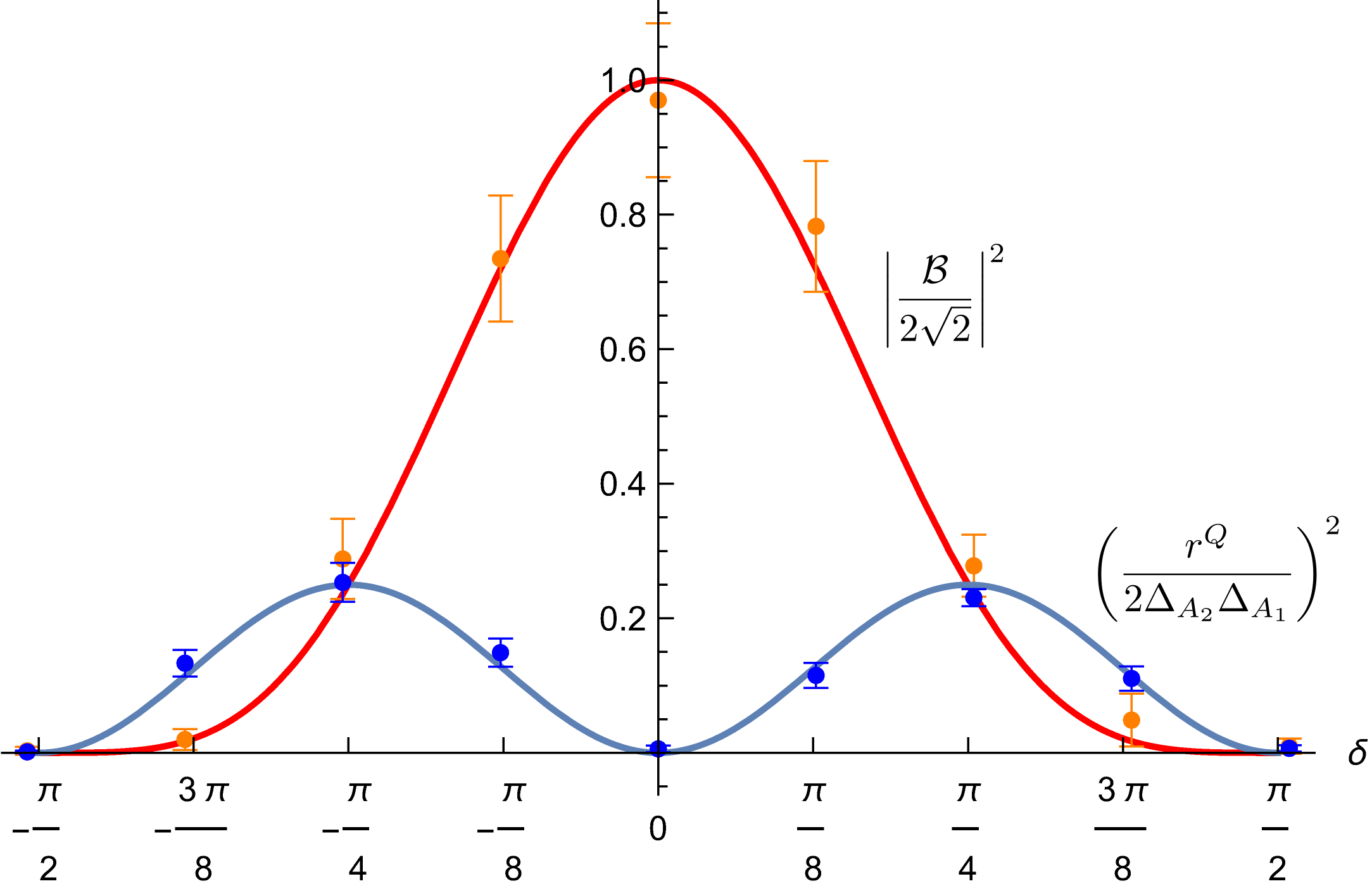}
		\subcaption{Separate contributions to $\mathcal{RI}$.}
		\label{sovrapposto}
	\end{subfigure}
		\caption{\textbf{Experimental test of the RI bound}.
Plot (a): test of the nonlocal correlations bound in Eq. \eqref{eq:RI-split}, derived from the Relativistic Independence principle, on the measurements performed by Alice and Bob on an entangled photon pair.
Specifically, the $\mathcal{RI}$ parameter is evaluated with Alice measuring the polarization of photon $A$ first in the $\{0,\frac{\pi}{2}\}$ basis and then in the $\{\frac{\pi}{4}+\delta,\frac{3\pi}{4}+\delta\}$ basis, with $\delta$ playing the role of a mismatch parameter with respect to the measurements allowing for a maximal violation of the CHSH inequality, while Bob realizes subsequent polarization measurements on photon $B$ in the $\{\frac{\pi}{8},\frac{5\pi}{8}\}$ and $\{\frac{3\pi}{8}+\delta,\frac{7\pi}{8}+\delta\}$ bases, respectively.
\newnew{Plot (b): interplay between local and non-local correlations within the bipartite system analyzed, highlighted by the behavior, w.r.t. the mismatch parameter $\delta$, of the two contributions to $\mathcal{RI}$, i.e.: the Tsirelson-bound-normalized Bell-CHSH parameter (red curve and orange dots), accounting for the non-local correlations; the Pearson correlation coefficient between the two measurements $A_1,A_2$ (azure curve and blue dots), accounting for the local correlations in Alice's lab.}
In both plots, dots and continuous curves represent the experimental results and the theoretical expectations, respectively.
Uncertainty bars account for statistical uncertainties and all other uncertainty contributions (see Appendix D for details).}
\label{results}
\end{figure}

The experimental results are shown in Fig. \ref{results}.
%
There, we plot the results obtained for the RI bound in Eq. \eqref{eq:RI-split} by choosing on Alice's side the $\{0,\frac{\pi}{2}\}$ measurement basis (being $\{\gamma,\gamma+\frac{\pi}{2}\}$ a basis defined by the rotation angle $\gamma$ with respect to the $\{H,V\}$ basis) on the first measurement block and the $\{\frac{\pi}{4}+\delta,\frac{3\pi}{4}+\delta\}$ basis on the second one, while Bob performs the first and second measurements in his lab in the $\{\frac{\pi}{8},\frac{5\pi}{8}\}$ and $\{\frac{3\pi}{8}+\delta,\frac{7\pi}{8}+\delta\}$ bases, respectively.
\new{We chose these measurement bases because, for $\delta=0$, they grant a maximal violation of the CHSH inequality, saturating the Tsirelson bound and maximizing the nonlocal correlations in our experiment (cancelling the local ones), as per Eq. \eqref{eq:RI-split}.
This means that, in our setup, the parameter $\delta$ plays the role of an ``angular mismatch'' in the second measurement of each observer with respect to the one maximizing the Bell-CHSH parameter $\left|\mathcal B\right|$; by varying $\delta$ we are able to quantitatively investigate the interplay between local and nonlocal correlations in our system.\\
Fig. \ref{results}a shows the $\mathcal{RI}$ parameter, quantifying the total contributions given by the local correlations on Alice's lab and the nonlocal correlations between Alice's and Bob's measurements, as a function of $\delta$.}
The experimental points (red dots) are in good agreement with the theoretical predictions (blue curve) within the experimental uncertainties (red vertical bars), and highlight how the nonlocal bound in Eq. \eqref{eq:RI-split} is always satisfied by the measurement correlations registered.
In plot (b), instead, we show the interplay between the two $\mathcal{RI}$ addenda, respectively accounting for nonlocal correlations of $A$ and $B$ (reddish theoretical curve and experimental dots) and the local correlations on Alice's side (blue theoretical curve and experimental dots).
In particular we observe how, for measurement choices closer to the ones allowing for a maximal violation of the CHSH inequality (i.e., for $\delta\in\left[-\frac\pi4,\frac\pi4\right]$), there is a trade-off between local and nonlocal correlations, the latter reaching their maximum for $\delta=0$, in correspondence with the RI bound saturation and in complete absence of local correlations.\\
\new{For this specific setting, our experiment yields $\mathcal{RI}=0.98\pm0.11$, in excellent agreement with the theoretical predictions just like all other points reported in Fig. \ref{totale}, further confirmed by looking separately at the local and nonlocal terms of Eq. \eqref{eq:RI-split} shown in Fig. \ref{sovrapposto}.
Table \ref{RI0table} reports the different contributions to the experimental uncertainty obtained for $\mathcal{RI}(\delta=0)$.}
\begin{table}[htbp]
	\centering
\begin{tabular}{|c|c|c|c|c|}
		\hline
		$\;\;\delta\;\;$ & $\;\;\mathcal{RI}\;\;$ & $\;\;\sigma_{\mathcal{RI}}\;\;$ & $\;\;\sigma_{\mathcal{RI}, \mathrm{stat}}\;\;$ & $\;\;\sigma_{\mathcal{RI}, \mathrm{cal}}\;\;$ \\
		\hline
		$\;\;0\;\;$ & $\;\;0.98\;\;$ & $\;\;0.11\;\;$ & $\;\;0.08\;\;$ & $\;\;0.08\;\;$ \\
		\hline
	\end{tabular}
\caption{\new{Uncertainty budget for the average value of the $\mathcal{RI}$ quantity estimated in our experiment for $\delta=0$, i.e. in correspondence to the RI bound saturation. The contributions to the overall experimental uncertainty $\sigma_\mathcal{RI}$ are (see Appendix D for details): $\sigma_{\mathcal{RI}, \mathrm{stat}}$ - statistical uncertainty contribution; $\sigma_{\mathcal{RI}, \mathrm{cal}}$ - uncertainty contribution accounting for the setup calibration procedure.}}
	\label{RI0table}
\end{table}
\new{Specifically, $\sigma_{\mathcal{RI}, \mathrm{stat}}$ is the statistical uncertainty given by the data set collected during the experimental run, while $\sigma_{\mathcal{RI}, \mathrm{cal}}$ accounts for the uncertainties arising from the measurement apparatus calibration procedure (see Appendix D for details on the experimental uncertainties evaluation).}\\

\section{Discussion}
The interplay between local and nonlocal correlations is of major relevance not just for the investigation of QM foundations \cite{gen10}, but also for quantum technologies, given the role of nonclassical correlations in quantum imaging and sensing, quantum metrology, quantum information processing and computation.
\newnew{Our experiment allows realizing for the first time simultaneous joint and sequential measurements on the same quantum system without collapsing its wave function,} providing an elaborate characterization of the set of quantum correlations, as derived from the Relativistic Independence principle \cite{car19}, and explicitly demonstrating an interplay between the amount of local and nonlocal correlations available in a quantum system.
Moreover, our work substantiates the hypothesis \cite{shi83,shi86} that quantum mechanics is as nonlocal as it is, without violating causality, thanks to the existence of uncertainty.
This result, following a rich and flourishing research path \cite{bar05,gis09,cav11,aol12,rin16,gis20,woo22,mar23,wan23,cab23,pet23}, represents a significant step toward understanding nonlocality in our most fundamental theory of nature \cite{gen19}.\\
On the practical side, such a measurement capability provides the key to investigate novel applications of quantum theory combining both local and nonlocal correlations.
Prime among these are quantum computation and simulation, which employ entanglement and nonlocal correlations, but balance them with local correlations to obtain optimal outcomes \cite{joz03,gro09}.
Furthermore, the proposed technique enables to measure both local and nonlocal correlations on the \newnew{same quantum system without inducing major decoherence on its state; this, for instance, can help assessing the performance of near-intermediate scale quantum devices without interfering much with their operation.}\\

\section*{Appendix A: Relativistic Independence bound derivation}
\newnew{As one can deduce from Eq. \eqref{genrel}, Bob can nonlocally tamper with the generalized uncertainty relation encoded in that positive semi-definite matrix, since $r_j$ depends in general on Bob's choice $j$. The RI principle is the requirement for the generalized uncertainty relations to exist and remain independent of Bob (Alice) measurement choices, i.e. $r_j\equiv r$ (see Eq. \eqref{RImat}).
It has been shown \cite{car19} that the RI principle establishes a bound on local and shared correlations between Alice and Bob measurements.
By the Schur complement \cite{lam16} condition for positive-semidefiniteness, one can write Eq. \eqref{genrel} as:
\begin{equation}
\begin{split}
M^{-1} \pmb{\Lambda}_A M^{-1}& =
\begin{bmatrix}
1&\frac{r}{\Delta_{A_2}\Delta_{A_1}}\\
\frac{r^*}{\Delta_{A_2}\Delta_{A_1}}&1
\end{bmatrix}\succeq
\begin{bmatrix}
|\rho_{j2}|^2 & \rho_{j1}\rho_{j2} \\
 \rho^*_{j1}\rho^*_{j2}& |\rho_{j1}|^2
\end{bmatrix}
\\
\begin{bmatrix}
|\rho_{j2}|^2 & \rho_{j1}\rho_{j2} \\
 \rho^*_{j1}\rho^*_{j2}& |\rho_{j1}|^2
\end{bmatrix}&= \Delta^{-2}_{B_j} M^{-1}
\begin{bmatrix}
C^*(B_j,A_2)\\
C^*(B_j,A_1)
\end{bmatrix}
\begin{bmatrix}
C(B_j,A_2) & C(B_j,A_1)
\end{bmatrix} M^{-1},
\end{split}
\label{schur}
\end{equation}
%
being $M$ a diagonal matrix with non-vanishing terms (not changing the positivity, just needed for transforming covariance into correlation), $\Delta_{A_2}$ and $\Delta_{A_1}$ the local variances, and $\rho_{ji}=(E_{B_j A_i} - E_{B_j}E_{A_i})/\Delta_{B_j}\Delta_{A_i}$ the Pearson correlation coefficient between $A_i$ and $B_j$.
The positive-semidefiniteness condition for a generic $n\times n$ matrix $\Gamma$ implies that, for any vector $u_j\in \mathbb{C}^n$, one has $u_j\Gamma u_j^\dagger\geq0$ \cite{coh20}.
If we consider $n=2$ and $u_j= [(-1)^j,1]$, by applying this property to Eq. \eqref{schur} one obtains:
\begin{equation}
u_j\left(
\begin{bmatrix}
1&\frac{r}{\Delta_{A_1}\Delta_{A_0}}\\
\frac{r^*}{\Delta_{A_1}\Delta_{A_0}}&1
\end{bmatrix}-
\begin{bmatrix}
|\rho_{1j}|^2 & \rho_{0j}\rho_{1j} \\
 \rho^*_{0j}\rho^*_{1j}& |\rho_{0j}|^2
\end{bmatrix} \right) u_j^\dagger\succeq 0\;,
\end{equation}
implying:
\begin{equation}
2\left(1+(-1)^j \Re\left(\frac{r}{\Delta_{A_2}\Delta_{A_1}}\right)\right) \geq | \rho_{1j} + (-1)^j \rho_{2j} |^2\;.
\label{passmath}
\end{equation}
By taking the square root on both sides, summing over Bob's choices $j=1,2$ and using the triangle inequality $|s|+|t|\geq|s+t|$, Eq. \eqref{passmath} implies:
\begin{equation}
|\mathcal{B}|\leq \sqrt{2} \sum_{j=1,2} \sqrt{1+(-1)^j \left(\frac{r}{\Delta_{A_2}\Delta_{A_1}}\right)},
\end{equation}
being $\mathcal{B}= \langle \hat{A}_1\hat{B}_1\rangle +\langle \hat{A}_1\hat{B}_2\rangle+\langle \hat{A}_2\hat{B}_1\rangle-\langle \hat{A}_2\hat{B}_2\rangle$ the Bell-CHSH parameter.
Now, since $\sqrt{1-a}\leq 1- a/2$ for $a\in[0,1]$, the RI bound follows (upon squaring and rearranging):
\begin{equation}
0\leq\mathcal{RI}=\left|\frac{\mathcal{B}}{2\sqrt{2}}\right|^2 + \left(\Re\left[ \frac{r}{2\Delta_{A_2}\Delta_{A_1}} \right]\right)^2 \leq 1\;,
\end{equation}
which for the QM case ($r=r^Q$) becomes:
\begin{equation}
0\leq\mathcal{RI}=\left|\frac{\mathcal{B}}{2\sqrt{2}}\right|^2 + \left(\frac{r^Q}{2\Delta_{A_2}\Delta_{A_1}}\right)^2 \leq 1\;.
\end{equation}
}
\\
\section*{Appendix B: experimental details}

\begin{figure}
\centering
\includegraphics[scale=1.3]{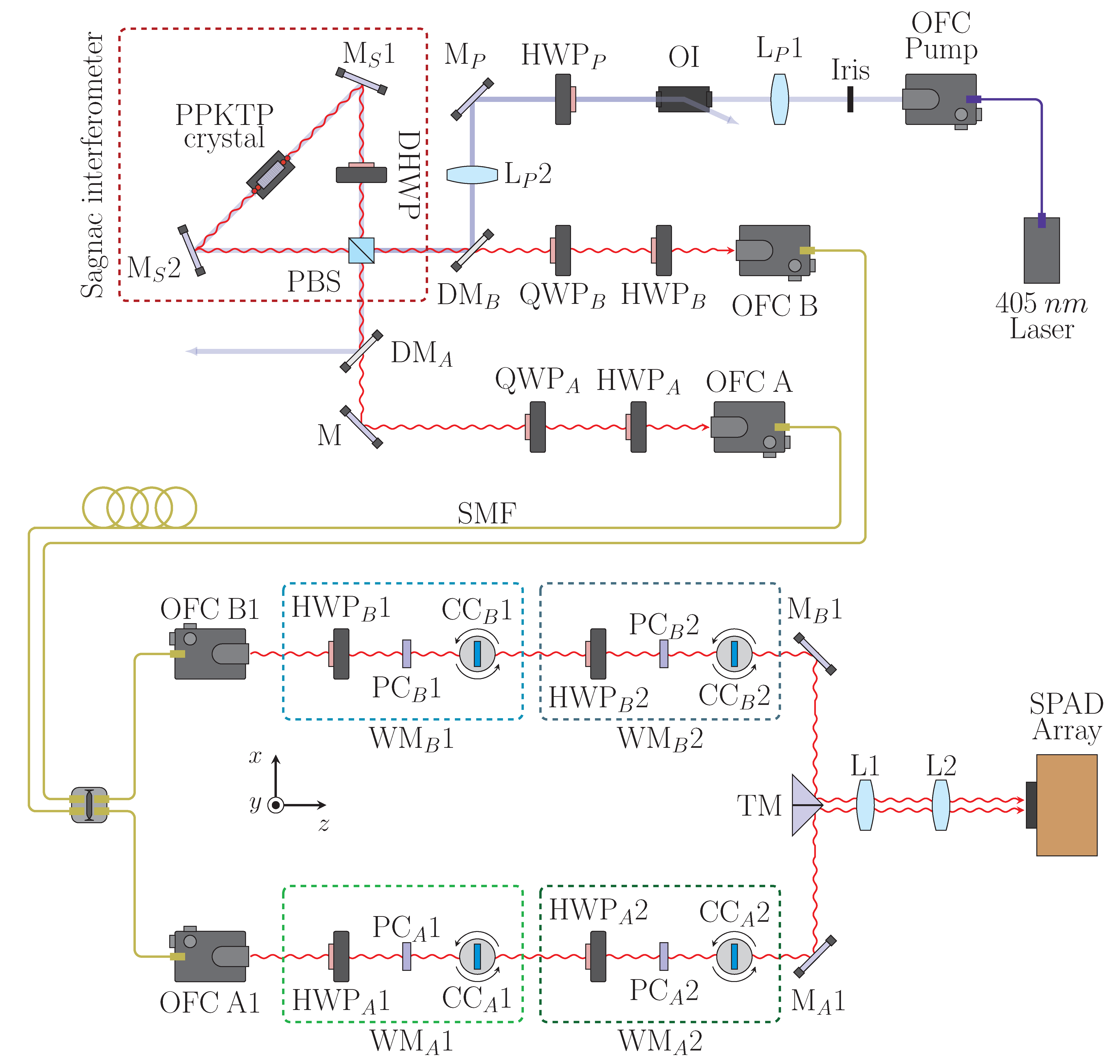}
\caption{Representation of our experimental setup. Polarization-entangled photon pairs in the singlet state $\ket{\psi_-}=(\ket{H_A V_B} -\ket{V_A H_B})/\sqrt{2}$ are generated in a Sagnac interferometer, coupled to an optical fiber and collimated in a Gaussian spatial mode.
Then, each photon undergoes two weak measurements in a row, each realized by a pair of birefringent crystals preceded by a half-wave plate. At the end of the measurement process, the photon pair is detected by a $24\times 24$ SPAD array with 2D spatial resolution. \newnew{CC: compensation crystal; DHWP: dual half-wave plate; DM: dichroic mirror; HWP: half-wave plate; Iris: pump beam pinhole; L1,L2: imaging system lenses; L$_p1$,L$_p2$: pump beam focusing lenses; M: mirror; OFC: optical fiber coupler; OI: optical isolator; PBS: polarizing beam splitter; PC: principal crystal; PPKTP: periodically-poled potassium titanyl phosphate; QWP: quarter-wave plate; SMF: single-mode fiber; SPAD: single-photon avalanche diode; TM: triangular mirror; WM: weak measurement.}}
\label{expapp}
\end{figure}

A detailed scheme of our experimental setup is presented in Fig. \ref{expapp}.
The setup consists of two main parts, one dedicated to the generation of polarization-entangled photon pairs, and the other focused on their (weak) measurement and detection.
The polarization-entangled two-photon state is produced by means of type-II spontaneous parametric down-conversion (SPDC) occurring in a periodically-poled potassium titanyl phosphate (PPKTP) crystal hosted in a Sagnac interferometer.
The PPKTP crystal is pumped by a CW laser at 405 nm, producing degenerate SPDC photon pairs at 810 nm.
The down-converted photons, exiting the Sagnac interferometer in the singlet state $\ket{\psi_-}=(\ket{H_A V_B} -\ket{V_A H_B})/\sqrt{2}$, are then spectrally filtered, coupled to single-mode fibers and collimated into Gaussian spatial modes $|f_{x_{A(B)}}\rangle\otimes|f_{y_{A(B)}}\rangle$ (with $\bra{\zeta}\ket{f_\zeta} =\frac{1}{(2\pi\sigma^2)^{1/4}}\exp\left(-\frac{\zeta^2}{4\sigma^2}\right)$, being $\zeta=x,y$ and $\sigma$ the Gaussian distribution width in both the $x$ and $y$ directions).
The combination of both quarter- and half-wave plates (QWP$_{A(B)}$ and HWP$_{A(B)}$, respectively) allows compensating for the polarization changes due to the propagation within the single-mode fibers.
This way, we are able to send photon pairs in the state $\ket{\Psi_{in}}= \ket{\psi_-} \otimes \ket{f_{x_A}} \otimes \ket{f_{y_A}} \otimes\ket{f_{x_B}} \otimes\ket{f_{y_B}}$ to the $A$ and $B$ branches with visibility $V^{in}=\frac{N(+,-)+N(-,+)-N(+,+)-N(-,-)}{N(+,-)+N(-,+)+N(+,+)+N(-,-)}=0.983\pm 0.001$, where $N(\alpha,\beta)$ is the number of two-photon counts registered while projecting photons $A$ and $B$ onto the $|\alpha\rangle$ and $|\beta\rangle$ states, respectively, and $\ket{\pm}=\frac{1}{\sqrt2}\left(\ket{H}\pm\ket{V}\right)$.\\
\new{The subsequent WMs are carried out by exploiting the weak coupling between polarization and transverse momentum induced on each photon by pairs of thin calcite (CaCO$_3$) birefringent crystals.}
Each pair is composed of a principal crystal (PC$_{A(B)}j$) and a compensation crystal (CC$_{A(B)}j$), where $j=1,2$ indicates the position of the crystal pair in the measurement sequence.
The optical $e$-axes of PC$_{A(B)}1$ and PC$_{A(B)}2$ are oriented along perpendicular planes, specifically the $z-x$ and $y-z$ planes (with a $\pi/4$ angle with respect to the photons propagation direction $z$), in order to induce a spatial walk-off (for the polarization component lying in the same plane as the crystal $e$-axis) in the two independent transverse directions $x$ and $y$, to avoid affecting each other.
Moreover, the PC$_{A(B)}j$ crystals also introduce a temporal delay and a phase shift between the $H$ and $V$ polarization components, which need to be compensated.
The compensation crystals CC$_{A(B)}1$ and CC$_{A(B)}2$ are mounted onto a motorized rotating stage, with the optical $e$-axis lying, respectively, in the $y$ and $x$ directions (i.e., orthogonal to both the $z$ direction and the principal crystal optical $e$-axis).
By rotating the compensation crystal along its optical axis, it becomes possible to restore the temporal delay and correct the phase without introducing any additional spatial decoherence.
\new{Each crystal pair is preceded by a half-wave plate (HWP$_{A(B)}$1 and HWP$_{A(B)}$2), enabling the choice of different polarization bases for each weak measurement implemented and allowing to realize the unitary transformation in Eq. \eqref{Ugen}.
Finally, entangled photons are detected by a 24$\times$24 Single Photon Avalanche Diode Array (SPADA), i.e. a single photon detector with 2D spatial resolution composed of an array of SPADs with internal time-tagging electronics \cite{mad21} able to record the arrival time (with 2 ns resolution) and position of each detected photon pair.
The L1 and L2 lenses constitute an imaging system needed to match the photon spatial distributions dimensions with the ones of the SPADA active area.
The internal time-tagger of the SPADA allows extracting the two-fold coincidences and the coordinates of the firing pixels, constituting the coincidence counts tensor $N(X_A,Y_A,X_B,Y_B)$, being $X_A$, $Y_A$ and $X_B$, $Y_B$ the pixel coordinates in which the two photons of each pair are detected.

\section*{Appendix C: Analysis of the state decoherence induced by the weak interactions}
After the four (weak) von Neumann couplings occurring in our setup (Fig. \ref{scheme}), the output state reads:
\begin{equation}
\ket{\Psi_{out}}=\hat{U}_{y_B}\hat{U}_{y_A}\hat{U}_{x_B}\hat{U}_{x_A}\ket{\Psi_{in}}=\hat{U}_{y_B}\hat{U}_{y_A}\hat{U}_{x_B}\hat{U}_{x_A}\ket{\psi_-}\otimes \ket{f_{x_A}} \otimes \ket{f_{y_A}} \otimes\ket{f_{x_B}} \otimes\ket{f_{y_B}}\;,
\end{equation}
where the $\hat{U}_{\zeta_{jK}}$'s ($j=1,2$, $K=A,B$) are the evolution operators defined in Eq. \eqref{Ugen}, $\ket{\psi_-}$ is the polarization-entangled singlet state generated by our EPS, and the $\ket{f_{\zeta_{jK}}}$'s are the (Gaussian) spatial components of the entangled pair wave function.
\newnew{This means that the overall density matrix of the (pure) output state can be written as $\rho_{out}=\ket{\Psi_{out}} \bra{\Psi_{out}}$.
If we restrict ourselves to the polarization subspace, the partial trace over the spatial degrees of freedom leads to the polarization state density matrix $\rho_{out}^{(\psi)}=\Tr_{x_A,y_A,x_B,y_B}\left( \rho_{out} \right)$, which presents some slight decoherence (due to the loss of information on the spatial degrees of freedom caused by the trace operation) with respect to the initial state $\rho_{in}=\ket{\Psi_{in}} \bra{\Psi_{in}}$ because of the occurring WMs.
We quantify this decoherence by comparing the purities $\mathcal{P}(\rho_{in}^{(\psi)})$ and $\mathcal{P}(\rho_{out}^{(\psi)})$ (being $\mathcal{P}(\rho)=\mathrm{Tr}(\rho^2)$ and $\rho_{in}^{(\psi)}=\Tr_{x_A,y_A,x_B,y_B}\left( \rho_{in} \right)$) of the polarization state density matrix before and after the WMs, both extracted via quantum tomographic reconstruction \cite{bog10}.
For the sake of simplicity, let us now assume that all the weak couplings in our setup have approximately the same strength, i.e., $g_{\zeta_{j K}}\simeq g$, and consequently let us introduce the parameter $\Omega=1-\exp\left( -\frac{g^2}{8\sigma^2} \right)$, playing the role of some ``decoherence parameter'' \cite{reb21a,reb21b} (as we will see in the following).
Given our weak interaction regime, in which $\frac{g}{\sigma}\ll1$ implies $\Omega\rightarrow0$, the output state purity $\mathcal{P}(\rho_{out}^{(\psi)})$ can be expanded around $\Omega=0$ as follows:
\begin{equation}
\begin{split}
\mathcal{P}(\rho_{out}^{(\psi)})\approx1&-4\Omega+\frac{1}{2} \Omega^2\Big(22+\cos[4(\alpha_1- \alpha_2)]+\cos[4(\alpha_1- \beta_1)]+\cos[4(\alpha_2- \beta_1)]+\cos[4(\alpha_1- \beta_2)] \\&+\cos[4(\alpha_2- \beta_2)]+\cos[4(\beta_1- \beta_2)]\Big)+ O(\Omega^3).
\end{split}
\label{pur}
\end{equation}
Remarkably, for the small interaction strength of the WMs implemented in our experiment ($\frac{g}{\sigma}\lesssim0.2$, corresponding to $\Omega\lesssim0.005$, a clear evidence of the weakness of our measurements), the angular dependence of Eq. \eqref{pur} becomes negligible, and the decoherence amount in $\rho_{out}^{(\psi)}$ is the same for every $\alpha_i,\beta_j$ choice.\\}
Since the singlet state produced by our EPS presents an initial purity of $0.98$, the induced decoherence should lead to an output state purity $\mathcal{P}(\rho_{out}^{(\psi)})\simeq0.96$.
The tomographically-reconstructed output state for $\delta=0$ featured, instead, $\mathcal{P}(\rho_{out}^{(\psi)})=0.94$, quite close to the theoretical expectations, with the further purity drop most likely due to the non-optimality of our optical components (not included in the theoretical analysis).
Nevertheless, these results demonstrate how, after the WM process, the polarization state has just suffered a slight decoherence, instead of the full wave function collapse eventually induced by strong (projective) measurements.
}

\section*{Appendix D: statistical analysis}

In our experiment, we perform six data acquisitions, with different setup configurations.
To calibrate the system, we send only either horizontally ($H$) or vertically ($V$) polarized photons to both $A$ and $B$, by generating the two-photon separable states $\ket{H_AV_B}$ and $\ket{V_AH_B}$.
\new{Considering that the first birefringent crystal pairs in $A$  and $B$ (respectively composed of the PC$_K1$ and CC$_K1$ crystals in Fig. \ref{expapp}, being $K=A,B$) shift the $H$-polarized photons along the $x$ transverse direction, while the second crystal pairs (formed by the PC$_K2$ and CC$_K2$ crystals) shift the $V$-polarized photons along $y$, the $\ket{H_AV_B}$ and $\ket{V_AH_B}$ states allow obtaining (via a linear regression of several subsets and subsequent average) the spatial distribution centres for the $H$ and $V$ photons.
From these acquisitions we can extract the photon distribution centres also in the unperturbed case and, as a consequence, the $g_{\zeta_{j K}}$ values.\\
For convenience, the centre of the unperturbed (shifted) spatial distribution is dubbed $\tilde{\zeta}_{K, 0(1)}$ ($\zeta = x, y$; $K = A, B$).
In light of this, the WM interaction intensities can be obtained as (see \cite{kof12,tam13} for further detail on the connection between spatial distributions and WM interaction intensity):
\begin{equation}\label{g_zeta}
  g_{\zeta_K} = \tilde{\zeta}_{K, 1} - \tilde{\zeta}_{K, 0}\;.
\end{equation}
Then, we set HWP$_P$ to produce the singlet state $|\psi_-\rangle$, and the HWPs on $A$ (HWP$_A1$ and HWP$_A2$) and $B$ (HWP$_B1$ and HWP$_B2$) sides so as to get a full data set allowing to evaluate the $\mathcal{RI}$ bound for a fixed ``mismatch parameter'' $\delta$, obtaining a set of $\zeta_{K}$ ($\zeta = x, y$; $K = A, B$) values.
Finally, we remove the PC$_Km$ and CC$_Km$ birefringent crystals ($m=1,2$) and perform three data acquisitions with only HWPs in the photon paths.
In the first acquisition, we keep the same HWP$_Km$'s angles as the $\mathcal{RI}$ acquisition, and send the singlet state $\ket*{\psi_-}$ as input.
Conversely, the second and third acquisitions are performed with all the HWP$_Km$'s on their 0's and, respectively, $\ket{H_AV_B}$ and $\ket{V_AH_B}$ as input states.}
This allows evaluating (and eventually compensating for) the unwanted spatial deviation induced on the photons by the HWP position change between the calibration and $\mathcal{RI}$ acquisitions, which we dub as $\tilde{\zeta}_{K, \mathrm{shift}}$.
Once again, they are extracted by linear regression and subsequent average.\\
\new{The left-hand side of the Relativistic Independence bound in Eq. \eqref{eq:RI-split} is the square sum of two terms, one ($\left|\frac{\mathcal B}{2\sqrt2}\right|$) concerning the nonlocal correlations between Alice and Bob and the other ($\Delta$) accounting for the local correlations on Alice's side.
Following the same mathematical derivation leading to Eq. \eqref{eq:RI-split}, we can express the Bell-CHSH parameter $\mathcal{B}$ as (from now on, we drop the $_{out}$ subscript for the sake of readability):}
%
\begin{align}
	\mathcal{B} = 4 \left( \frac{\ev*{\hat{X}_A \hat{X}_B}}{g_{X_A} g_{X_B}} - \frac{\ev*{\hat{X}_A \hat{Y}_B}}{g_{X_A} g_{Y_B}} + \frac{\ev*{\hat{Y}_A \hat{X}_B}}{g_{Y_A} g_{X_B}} + \frac{\ev*{\hat{Y}_A \hat{Y}_B}}{g_{Y_A} g_{Y_B}} - \frac{\ev*{\hat{Y}_A}}{g_{Y_A}} - \frac{\ev*{\hat{X}_B}}{g_{X_B}} \right) +2
	\label{eq:b-explicit}
\end{align}
The expressions $\ev*{\hat{\zeta}_{K}}$ ($\zeta = X, Y$; $K = A, B$) and $\ev{\hat{\zeta}_{A} \otimes \hat{\zeta}_{B}}$ ($\zeta = X, Y$; $K = A, B$) can be evaluated as
\begin{align}
	\ev{\hat{\zeta}_{K}} &= \ev{\zeta_{K} - \tilde{\zeta}_{K, 0} - \tilde{\zeta}_{K,\mathrm{shift}}}\\
	\ev{\hat{\zeta}_{A} \otimes \hat{\zeta}_{B}} &=  \ev{(\zeta_{A} - \tilde{\zeta}_{A, 0} - \tilde{\zeta}_{A,\mathrm{shift}})(\zeta_{B} - \tilde{\zeta}_{B, 0} - \tilde{\zeta}_{B,\mathrm{shift}})}\;,
	\label{eq:zeta-exp}
\end{align}
Substituting in Eq. \eqref{eq:b-explicit}, we obtain the following expression for the Bell-CHSH parameter:
\begin{align}
	\mathcal{B} = 4\bigg\langle & \frac{(X_{A} - \tilde{X}_{A, 0} - \tilde{X}_{A, \mathrm{shift}}) (X_{B} - \tilde{X}_{B, 0} - \tilde{X}_{B, \mathrm{shift}})}{(\tilde{X}_{A, 1} - \tilde{X}_{B, 0}) (\tilde{X}_{B, 1} - \tilde{X}_{B, 0})}\nonumber\\
	-& \frac{(X_{A} - \tilde{X}_{A, 0} - \tilde{X}_{A, \mathrm{shift}}) (Y_{B} - \tilde{Y}_{B, 0} - \tilde{Y}_{B, \mathrm{shift}})}{(\tilde{X}_{A, 1} - \tilde{X}_{A, 0}) (\tilde{Y}_{B, 1} - \tilde{Y}_{B, 0})}\nonumber\\
	+& \frac{(Y_{A} - \tilde{Y}_{A, 0} - \tilde{Y}_{A, \mathrm{shift}}) (X_{B} - \tilde{X}_{B, 0} - \tilde{X}_{B, \mathrm{shift}})}{(\tilde{Y}_{A, 1} - \tilde{y}_{A, 0}) (\tilde{X}_{B, 1} - \tilde{X}_{B, 0})}\nonumber\\
	+& \frac{(Y_{A} - \tilde{Y}_{A, 0} - \tilde{Y}_{A, \mathrm{shift}}) (Y_{B} - \tilde{Y}_{B, 0} - \tilde{Y}_{B, \mathrm{shift}})}{(\tilde{Y}_{A, 1} - \tilde{Y}_{A, 0}) (\tilde{Y}_{B, 1} - \tilde{Y}_{B, 0})}\nonumber\\
		-& \frac{Y_{A} - \tilde{Y}_{A, 0} - \tilde{Y}_{A, \mathrm{shift}}}{\tilde{Y}_{A, 1} - \tilde{Y}_{A, 0}}
		-\frac{X_{1B} - \tilde{X}_{B, 0} - \tilde{X}_{B, \mathrm{shift}}}{\tilde{X}_{B, 1} - \tilde{X}_{B, 0}}
		\bigg\rangle+2\;.
	\label{eq:b-operative}
\end{align}
Conversely, the $\Delta$ quantity in Eq. \eqref{eq:RI-split} can be expressed as:
\begin{align}
	\Delta &= \frac{\frac{\ev*{\hat{X}_A \hat{Y}_A}}{g_{X_A} g_{Y_A}} - \frac{\ev*{\hat{X}_A} \ev*{\hat{Y}_A}}{g_{X_A} g_{Y_A}}}{2 \sqrt{ \left( \frac{\ev*{\hat{X}_A}}{g_{X_A}} - \frac{\ev*{\hat{X}_A}^2}{g^2_{X_A}} \right) \left( \frac{\ev*{\hat{Y}_A}}{g_{Y_A}} - \frac{\ev*{\hat{Y}_A}^2}{g^2_{Y_A}} \right)}} =
	\frac{\ev*{\hat{X}_A \hat{Y}_A} - \ev*{\hat{X}_A} \ev*{\hat{Y}_A}}{2 \sqrt{ \left( g_{X_A} \ev*{\hat{X}_A} - \ev*{\hat{X}_A}^2 \right) \left( g_{Y_A} \ev*{\hat{Y}_A} - \ev*{\hat{Y}_A}^2 \right)}}\nonumber\\
	& = \frac{C_{xy, A}}{2 \sqrt{S_{x, A} S_{y, A}}}\;,
	\label{eq:s-explicit}
\end{align}
where
\begin{align}
C_{xy, A} &= \ev*{\hat{X}_A \hat{Y}_A} - \ev*{\hat{X}_A} \ev*{\hat{Y}_A} \label{eq:def-per-Cxy}\\
S_{\zeta, A} &= g_{\zeta_A} \ev*{\hat{\zeta}_A} - \ev*{\hat{\zeta}_A}^2
	\label{eq:def-per-s}
\end{align}
with $\zeta = x, y$ and $\hat{\zeta} = \hat{X},\hat{Y}$.
\newnew{Substituting $\hat{\zeta}_K = \zeta_K - \tilde{\zeta}_{K, 0} - \tilde{\zeta}_{K, \mathrm{shift}}$ into Eq. \eqref{eq:def-per-Cxy} one obtains:
\begin{align}
	C_{xy, A} &= \ev{ \left( X_A - \tilde{X}_{A, 0} - \tilde{X}_{A, \mathrm{shift}} \right) \left( Y_A - \tilde{Y}_{A, 0} - \tilde{Y}_{A, \mathrm{shift}} \right) }\nonumber\\
	 &- \ev*{ X_A - \tilde{X}_{A, 0} - \tilde{X}_{A, \mathrm{shift}}} \ev*{ Y_A - \tilde{X}_{A, 0} - \tilde{Y}_{A, \mathrm{shift}}}\nonumber\\
	 &= \ev*{{X}_A {Y}_A} - \ev*{{X}_A} \ev*{{Y}_A}
	\label{eq:c-lemma1}
\end{align}
which follows from linearity of the expectation value.
By comparing Eq. \eqref{eq:def-per-Cxy} with Eq. \eqref{eq:c-lemma1} it appears evident how $C_{xy, A}$ results independent of both the set of coordinates of the unperturbed distribution center $(\tilde{X}_{A, 0},\tilde{Y}_{A, 0})$ and the one accounting for the unwanted deviations induced by the HWPs $(\tilde{X}_{A, \mathrm{shift}},\tilde{Y}_{A, \mathrm{shift}})$.}\\
Furthermore, simple manipulation shows that:
\begin{equation}
	S_{\zeta, A} = \left( \ev*{\zeta_A} - \tilde{\zeta}_{A, 0} - \tilde{\zeta}_{A, \mathrm{shift}} \right) \left( \tilde{\zeta}_{A, 1} + \tilde{\zeta}_{A, \mathrm{shift}} - \ev*{\zeta_A} \right)\;,
	\label{eq:sx-lemma2}
\end{equation}
allowing to write $\Delta$ as:
\begin{align}
	\Delta &= \frac{\ev*{X_A Y_A} - \ev*{X_A} \ev*{Y_A}}{2 \sqrt{ \left( \ev*{X_A} - \tilde{X}_{A, 0} - \tilde{X}_{A, \mathrm{shift}} \right) \left( \tilde{X}_{A, 1} + \tilde{X}_{A, \mathrm{shift}} - \ev*{X_A} \right) }} \nonumber\\
	&\times \frac{1}{\sqrt{\left( \ev*{Y_A} - \tilde{Y}_{A, 0} - \tilde{Y}_{A, \mathrm{shift}} \right) \left( \tilde{Y}_{A, 1} + \tilde{Y}_{A, \mathrm{shift}} - \ev*{Y_A} \right)}}\;.
	\label{eq:s-operative1}
\end{align}

For each quantity $\mathcal{P} = \{ \mathcal{B}, \Delta, \mathcal{RI}\}$, the associated experimental uncertainty is evaluated as
\begin{align}
	\sigma_{\mathcal{P}} &= \sqrt{ \sigma^2_{\mathcal{P}, \mathrm{stat}} + \sigma^2_{\mathcal{P}, \mathrm{cal}} }\;,
	\label{eq:uncertainty}
\end{align}
where $\sigma_{\mathcal{P}, \mathrm{stat}}$ represents the statistical contribution to the overall uncertainty and
\begin{equation}
	\sigma_{\mathcal{P}, \mathrm{cal}} = \sqrt{\sum_{\substack{\tilde{\zeta}=\tilde{X},\tilde{Y}\\ K=A, B\\ j=0, 1}} \left( \pdv{\mathcal{P}}{\tilde{\zeta}_{K, j}} \right)^2 \sigma_{\mathcal{P},\tilde{\zeta}_{K, j}}^2 + \sum_{\substack{\tilde{\zeta}=\tilde{X},\tilde{Y}\\ K=A, B}} \left( \pdv{\mathcal{P}}{\tilde{\zeta}_{K, \mathrm{shift}}} \right)^2 \sigma_{\mathcal{P},\tilde{\zeta}_{K, \mathrm{shift}}}^2}
	\label{eq:uncertainty-cal}
\end{equation}
is the uncertainty associated with the calibration procedure (including the HWP-induced spatial shift), with $\sigma_{\mathcal{P},\tilde{\zeta}_{K, j}}$ being the uncertainty contribution to $\sigma_{\mathcal{P}, \mathrm{cal}}$ depending on the calibration parameters $\tilde{\zeta}_{K, j}$ and $\sigma_{\mathcal{P},\tilde{\zeta}_{K, \mathrm{shift}}}$ being the uncertainty contribution associated with the spatial shift induced by the HWPs.\\
To evaluate the statistical uncertainty $\sigma_{\mathcal{P}, \mathrm{stat}}$, we assume the following hypotheses:
\begin{itemize}
	\item Let $\zeta_{K,i}$, i.e. the position $\zeta$ of the $i$-th detected photon on the branch $K$, be a random variable with variance $V^2 \left[ \zeta_{K,i} \right] = V^2_{\zeta_K}$ (being $\zeta = X, Y$).
	\item For independent events, the covariance condition $\sigma_{X_{K,i} Y_{K^\prime,j}} = \sigma_{X_K Y_{K^\prime}} \delta_{ij}$ is satisfied.
\end{itemize}
Then, we can write $\sigma_{\mathcal{P}, \mathrm{stat}}$ as:
\begin{equation}
	\sigma_{\mathcal{P}, \mathrm{stat}} = \sqrt{\sum_{\substack{\zeta=X, Y\\ K=A, B}} \left[
		\sum_{i=1}^N \left( \pdv{\mathcal{P}}{\zeta_{K,i}} \right)^2 \sigma^2_{\zeta_K} \right] +
		\sum_{\substack{\zeta=X, Y\\ K=A, B\\ K^\prime \neq K =A, B}} \left[
		\sum_{j=1}^N \left( \pdv{\mathcal{P}}{\zeta_{K,j}} \right) \left( \pdv{\mathcal{P}}{\zeta_{K^\prime,j}} \right) \sigma_{\zeta_K \zeta_{K^\prime}} \right]}
\label{eq:uncertainty-stat}
\end{equation}
where $N$ is the total number of detected photon.
Note that, for $\mathcal{P} = \Delta$, only terms with $K=A$ contribute to the uncertainty.\\
\newnew{Furthermore, in order to quantify separately the uncertainty contribution associated with the local and non-local part of the $\mathcal{RI}$ quantity in Eq. \eqref{eq:RI-split}, we rewrite it as:
\begin{equation}\label{eq:RI-split_unc}
  \mathcal{RI}=\abs{\frac{\mathcal{B}}{2\sqrt{2}}}^2 + \Delta^2\equiv \mathcal{RI_B}+ \mathcal{RI}_\Delta\;,
\end{equation}
with $\mathcal{RI_B}=\abs{\frac{\mathcal{B}}{2\sqrt{2}}}^2$ and $\mathcal{RI}_\Delta=\Delta^2$ accounting, respectively, for the non-local and local correlations contribution to $\mathcal{RI}$.
By exploiting Eqs. \eqref{eq:uncertainty}-\eqref{eq:uncertainty-stat} with $\mathcal{P}=\mathcal{RI_B},\mathcal{RI}_\Delta$, we obtain the overall uncertainties $\sigma_{\mathcal{RI_B}}$ and $\sigma_{\mathcal{RI}_\Delta}$ associated with $\mathcal{RI_B}$ and $\mathcal{RI}_\Delta$, achieving a good estimate of the $\sigma_{\mathcal{RI}}$ amount due to non-local and local correlations (although the intertwining between local and non-local terms in $\sigma_{\mathcal{RI}}$ makes it impossible to write it as $\sigma_{\mathcal{RI}}=\sqrt{\sigma_{\mathcal{RI_B}}^2+\sigma_{\mathcal{RI}_\Delta}^2}$).\\
A detailed uncertainty budget is reported in Tables \ref{tab:budget-B}, \ref{tab:budget-S} and \ref{tab:budget}.}
\begin{table}[h!]
	\centering
\begin{tabular}{|c|c|c|c|c|}
		\hline
		$\;\;\delta\;\;$ & $\;\;\mathcal{B}\;\;$ & $\;\;\sigma_{\mathcal{B}}\;\;$ & $\;\;\sigma_{\mathcal{B}, \mathrm{stat}}\;\;$ & $\;\;\sigma_{\mathcal{B}, \mathrm{cal}}\;\;$ \\
		\hline
		$\;\;-\pi /2 \;\; $ & $\;\;-0.15\;\;$ & $\;\;0.16\;\;$ & $\;\;0.11\;\;$ & $\;\;0.11\;\;$ \\
		$\;\;-3\pi / 8\;\; $ & $\;\;-0.40\;\;$ & $\;\;0.15\;\;$ & $\;\;0.11\;\;$ & $\;\;0.11\;\;$ \\
		$\;\;-\pi / 4\;\; $ & $\;\;-1.52\;\;$ & $\;\;0.16\;\;$ & $\;\;0.12\;\;$ & $\;\;0.10\;\;$ \\
		$\;\;-\pi / 8\;\; $ & $\;\;-2.42\;\;$ & $\;\;0.15\;\;$ & $\;\;0.11\;\;$ & $\;\;0.10\;\;$ \\
		$\;\;0\;\;$ & $\;\;-2.79\;\;$ & $\;\;0.16\;\;$ & $\;\;0.12\;\;$ & $\;\;0.11\;\;$  \\
		$\;\;\pi / 8 \;\;$ & $\;\;-2.50\;\;$ & $\;\;0.16\;\;$ & $\;\;0.11\;\;$ & $\;\;0.11\;\;$ \\
		$\;\;\pi / 4 \;\;$ & $\;\;-1.49\;\;$ & $\;\;0.12\;\;$ & $\;\;0.05\;\;$ & $\;\;0.11\;\;$ \\
		$\;\;3\pi / 8\;\; $ & $\;\;-0.63\;\;$ & $\;\;0.25\;\;$ & $\;\;0.12\;\;$ & $\;\;0.22\;\;$ \\
		$\;\;\pi / 2\;\; $ & $\;\;-0.28\;\;$ & $\;\;0.16\;\;$ & $\;\;0.12\;\;$ & $\;\;0.11\;\;$ \\
		\hline
\end{tabular}
\caption{Uncertainty budget for the average value of the Bell-CHSH parameter $\mathcal{B}$ estimated in our experiment. $\sigma_\mathcal{B}$: total uncertainty. $\sigma_{\mathcal{B}, \mathrm{stat}}$: statistical uncertainty contribution. $\sigma_{\mathcal{B}, \mathrm{cal}}$: uncertainty contribution due to the setup calibration procedure and the correction of the HWP-related (unwanted) spatial shifts.}
\label{tab:budget-B}
\end{table}

\begin{table}[h!]
	\centering
\begin{tabular}{|c|c|c|c|c|}
		\hline
		$\;\;\delta\;\;$ & $\;\;\Delta\;\;$ & $\;\;\sigma_{\Delta}\;\;$ & $\;\;\sigma_{\Delta, \mathrm{stat}}\;\;$ & $\;\;\sigma_{\Delta, \mathrm{cal}}\;\;$ \\
		\hline
		$\;\;-\pi / 2 \;\;$ & $\;\;-0.037\;\;$ & $\;\;0.027\;\;$ & $\;\;0.027\;\;$ & $\;\;6.6 \times 10^{-5}\;\;$ \\
		$\;\;-3 \pi / 8\;\; $ & $\;\;-0.365\;\;$ & $\;\;0.027\;\;$ & $\;\;0.027\;\;$ & $\;\;5.6 \times 10^{-4}\;\;$ \\
		$\;\;-\pi / 4\;\; $ & $\;\;-0.503\;\;$ & $\;\;0.029\;\;$ & $\;\;0.029\;\;$ & $\;\;1.1 \times 10^{-3}\;\;$ \\
		$\;\;-\pi / 8 \;\;$ & $\;\;-0.386\;\;$ & $\;\;0.027\;\;$ & $\;\;0.027\;\;$ & $\;\;5.0 \times 10^{-4}\;\;$ \\
		$\;\;0\;\;$ & $\;\;-0.076\;\;$ & $\;\;0.031\;\;$ & $\;\;0.031\;\;$ & $\;\;1.7 \times 10^{-3}\;\;$ \\
		$\;\;\pi / 8 \;\;$ & $\;\;0.339\;\;$ & $\;\;0.027\;\;$ & $\;\;0.027\;\;$ & $\;\;4.3 \times 10^{-4}\;\;$ \\
		$\;\;\pi / 4 \;\;$ & $\;\;0.480\;\;$ & $\;\;0.013\;\;$ & $\;\;0.013\;\;$ & $\;\;9.2 \times 10^{-4}\;\;$ \\
		$\;\;3 \pi / 8 \;\;$ & $\;\;0.333\;\;$ & $\;\;0.027\;\;$ & $\;\;0.027\;\;$ & $\;\;6.4 \times 10^{-4}\;\;$ \\
		$\;\;\pi / 2 \;\;$ & $\;\;-0.081\;\;$ & $\;\;0.030\;\;$ & $\;\;0.030\;\;$ & $\;\;1.7 \times 10^{-4}\;\;$ \\
		\hline
\end{tabular}
\caption{Uncertainty budget for Alice's local correlations parameter $\Delta$ estimated in our experiment. $\sigma_\Delta$: total uncertainty. $\sigma_{\Delta, \mathrm{stat}}$: statistical uncertainty contribution. $\sigma_{\Delta, \mathrm{cal}}$: uncertainty contribution due to the setup calibration procedure and the HWP-related (unwanted) spatial shifts.}
	\label{tab:budget-S}
\end{table}

\begin{table}[h!]
	\centering
\begin{tabular}{|c|c|c|c|c|c|c|c|c|}
		\hline
		$\;\;\delta\;\;$ & $\;\;\mathcal{RI}\;\;$ & $\;\;\sigma_{\mathcal{RI}}\;\;$ & $\;\;\sigma_{\mathcal{RI}, \mathrm{stat}}\;\;$ & $\;\;\sigma_{\mathcal{RI}, \mathrm{cal}}\;\;$ & $\;\;\mathcal{RI_B}\;\;$ & $\;\;\sigma_{\mathcal{RI_B}}\;\;$ &$\mathcal{RI}_{\Delta}$ & $\;\;\sigma_{\mathcal{RI}_{\Delta}}$\;\; \\
		\hline
	
		$\;\;-\pi/ 2\;\;  $ & $\;\;0.0043\;\;$ & $\;\;0.0056\;\;$ & $\;\;0.0037\;\;$ & $\;\;0.0042\;\;$ & $\;\;0.0029\;\;$ & $\;\;0.0060\;\;$ & $\;\;0.0014\;\;$ & $\;\;0.0020\;\;$ \\
		
		$\;\;-3\pi / 8\;\; $ & $\;\;0.153\;\;$ & $\;\;0.024\;\;$ & $\;\;0.022\;\;$ & $\;\;0.010\;\;$ & $\;\;0.020\;\;$ & $\;\;0.015\;\;$ & $\;\;0.133\;\;$ & $\;\;0.020\;\;$ \\
	
		$\;\;-\pi / 4\;\; $ & $\;\;0.542\;\;$ & $\;\;0.059\;\;$ & $\;\;0.044\;\;$ & $\;\;0.039\;\;$ & $\;\;0.288\;\;$ & $\;\;0.059\;\;$ & $\;\;0.253\;\;$ & $\;\;0.029\;\;$ \\
	
		$\;\;-\pi / 8\;\; $ & $\;\;0.884\;\;$ & $\;\;0.082\;\;$ & $\;\;0.054\;\;$ & $\;\;0.063\;\;$ & $\;\;0.735\;\;$ & $\;\;0.094\;\;$ & $\;\;0.149\;\;$ & $\;\;0.021\;\;$ \\

		$\;\;0\;\;$ & $\;\;0.98\;\;$ & $\;\;0.11\;\;$ & $\;\;0.085\;\;$ & $\;\;0.077\;\;$ & $\;\;0.97\;\;$ & $\;\;0.11\;\;$ & $\;\;0.0058\;\;$ & $\;\;0.0048\;\;$ \\

		$\;\;\pi/ 8\;\;  $ & $\;\;0.898\;\;$ & $\;\;0.086\;\;$ & $\;\;0.054\;\;$ & $\;\;0.067\;\;$ & $\;\;0.783\;\;$ & $\;\;0.097\;\;$ & $\;\;0.115\;\;$ & $\;\;0.018\;\;$ \\

		$\;\;\pi/ 4\;\;  $ & $\;\;0.509\;\;$ & $\;\;0.045\;\;$ & $\;\;0.019\;\;$ & $\;\;0.041\;\;$ & $\;\;0.278\;\;$ & $\;\;0.046\;\;$ & $\;\;0.231\;\;$ & $\;\;0.013\;\;$ \\

		$\;\;3\pi / 8\;\; $ & $\;\;0.159\;\;$ & $\;\;0.041\;\;$ & $\;\;0.022\;\;$ & $\;\;0.035\;\;$ & $\;\;0.049\;\;$ & $\;\;0.039\;\;$ & $\;\;0.111\;\;$ & $\;\;0.018\;\;$ \\

		$\;\;\pi / 2\;\; $ & $\;\;0.017\;\;$ & $\;\;0.011\;\;$ & $\;\;0.0078\;\;$ & $\;\;0.0075\;\;$ & $\;\;0.010\;\;$ & $\;\;0.011\;\;$ & $\;\;0.0066\;\;$ & $\;\;0.0048\;\;$ \\
		\hline
	\end{tabular}
\caption{\newnew{Uncertainty budget for the experimental estimation of the three elements presented in Eq. \eqref{eq:RI-split_unc}, i.e.: relativistic independence parameter $\mathcal{RI}$; $\mathcal{RI_B} = \abs{\mathcal{B} / 2\sqrt{2}}^2 $, accounting for the non-local correlations contribution to $\mathcal{RI}$; $\mathcal{RI}_{\Delta} = \Delta^2$, constituting the $\mathcal{RI}$ part due to local correlations. $\sigma_\mathcal{RI}$: overall uncertainty associated with $\mathcal{RI}$. $\sigma_{\mathcal{RI}, \mathrm{stat}}$: statistical uncertainty contribution to $\sigma_\mathcal{RI}$. $\sigma_{\mathcal{RI}, \mathrm{cal}}$: uncertainty contribution to $\sigma_\mathcal{RI}$ due to the setup calibration procedure and the correction of the HWP-related (unwanted) spatial shifts. $\sigma_\mathcal{RI_B}$: overall uncertainty associated with the $\mathcal{RI}$ non-local component $\mathcal{RI_B}$. $\sigma_{\mathcal{RI}_{\Delta}}$: overall uncertainty associated with the $\mathcal{RI}$ local component $\mathcal{RI}_{\Delta}$.}}
	\label{tab:budget}
\end{table}

\section*{Acknowledgements}
This work was financially supported by the projects QuaFuPhy (call ``Trapezio'' of Fondazione San Paolo) and AQuTE (MUR, call ``PRIN 2022'', grant No. 2022RATBS4), by the Israel Science Foundation (grant No. 2208/24), by the Israel Innovation Authority under grant 70002 and grant 73795, by the Pazy Foundation, the Israeli Ministry of Science and Technology, and by the Quantum Science and Technology Program of the Israeli Council of Higher Education.
This work was also funded by the project EMPIR 19NRM06 METISQ.
This project received funding by the EMPIR program cofinanced by the Participating States and from the European Union Horizon 2020 Research and Innovation Programme.
The results presented in this article had been achieved also in the context of the following projects: QUID (QUantum Italy Deployment) and EQUO (European QUantum ecOsystems), which are funded by the European Commission in the Digital Europe Programme under the grant agreements number 101091408 and 101091561; QU-TEST, which had received funding from the European Union's Horizon Europe under the grant agreement number 101113901.
We thank Yakir Aharonov, Avishy Carmi and Avshalom Elitzur for enlightening discussions.




%

\end{document}